\documentclass{andromedaone}       

\journal{NDM}%Letters in High Energy Physics}
\vol{2020}
\jyear{Egypt}
\pages{Hurghada} 

\received{xx January 2018}
\published{xx March 2018}

\def\be{\begin{equation}}
\def\ee{\end{equation}}
\def\bea{\begin{eqnarray}}
\def\eea{\end{eqnarray}}

\usepackage{amsmath}
\usepackage{amsfonts}
\usepackage{amssymb}
\usepackage{slashed}

\begin{document}

\title{Exploring neutrino physics through sneutrinos}

\author{Stefano Moretti,\auno{1,2} Claire Shepherd-Themistocleous,\auno{1,2} and Harri Waltari\auno{1,2,3}}
\address{$^{1}$Particle physics department, STFC/Rutherford Appleton Laboratory, Chilton, Didcot, OX11 0QX, United Kingdom}
\address{$^{2}$Physics \& Astronomy, University of Southampton, Southampton, SO17 1BJ, United Kingdom}
\address{$^3$Department of physics, University of Helsinki, P.O. box 64, 00014 Helsinki, Finland}

\begin{abstract}
Supersymmetry relates neutrinos with their superpartners, sneutrinos. Unlike
neutrinos, sneutrinos may decay visibly in colliders. We discuss how we could get
information from neutrino Yukawa couplings in the NMSSM extended with right-
handed neutrinos, if the right-handed sneutrinos are within the reach of the colliders.
\end{abstract}

\maketitle


\section{Introduction}

The Standard Model (SM) of particle physics has several shortcomings, neutrino mass generation and the lack for a dark matter candidate being among the most pressing ones. 

Neutrinos have nonzero masses since neutrino oscillations exist \cite{Athanassopoulos:1997pv,Fukuda:1998mi}. The masses are  at most at the eV-scale \cite{Aker:2019uuj} and even smaller, if we assume standard cosmology \cite{Aghanim:2018eyx}. It is possible to introduce neutrino masses by adding right-handed neutrinos and Yukawa couplings to the SM, but the masses are so tiny that it seems likely that they are generated in a different manner.

If one extends the SM with $d=5$ terms, one has the Weinberg operator \cite{Weinberg:1979sa}, which gives rise to Majorana masses for neutrinos. This operator may be a remnant of a seesaw mechanism, where the SM is extended by either right-handed neutrinos (type-I), a scalar triplet (type-II) or a fermionic triplet (type-III) after integrating out the new particles, which have to be reasonably heavy so that they would not have been detected by now.

Supersymmetry is a framework that offers solutions to a number of the SM problems, but the minimal extension does not have a mechanism for neutrino mass generation. Various neutrino mass generation mechanisms can be embedded in a supersymmetric context by adding new superfields and extending the superpotential.

Supersymmetry introduces superpartners to all SM particles, the superpartner of a neutrino is a sneutrino. Typically we do not expect the sneutrino to be the lightest of the superpartners so it will decay. If there is a lighter charged superpartner, the sneutrino may give visible signatures at colliders. The importance of this for neutrino physics is that due to supersymmetry, sneutrinos inherit the interactions of neutrinos and hence the observation of a sneutrino signal could give us information about the neutrino mass generation mechanism. Here we construct an explicit example of such a case based on our work \cite{Moretti:2019yln}.

\section{Sneutrino physics in NMSSM with right-handed neutrinos}

The next-to-minimal supersymmetric Standard Model with right-handed neutrinos solves two of the problems of the MSSM, namely neutrino mass generation and the $\mu$-problem. It is based on the superpotential \cite{Kitano:1999qb,Cerdeno:2008ep}
\begin{equation}\label{eq:superpotential}
W=y^{u}_{ij}(Q_{i}\cdot H_{u})U^{c}_{j}-y^{d}_{ij}(Q_{i}\cdot H_{d})D^{c}_{j}-y^{\ell}_{ij}(L_{i}\cdot H_{d})E^{c}_{j}+y^{\nu}_{ij}(L_{i}\cdot H_{u})N^{c}_{i}
+\lambda S(H_{u}\cdot H_{d})+\frac{\lambda_{Ni}}{2}SN_{i}^{c}N_{i}^{c}+\frac{\kappa}{3} S^{3}.
\end{equation}
After the scalar component of the singlet superfield $S$ gets a vacuum expectation value, mass terms for both the higgsinos and right-handed neutrinos are generated.

As this is an electroweak scale seesaw model, the neutrino Yukawa couplings need to be tiny ($<10^{-6}$) in order to have sub-eV scale masses for the neutrinos.

The scalar potential has a term
\begin{equation}\label{eq:scalarpotential}
V=|\lambda H_{u}H_{d}+\lambda_{N} \tilde{N}^{2} +\kappa S^{2} |^{2}+\ldots
\end{equation}
The cross terms generate a lepton number violating mass term for right-handed sneutrinos after electroweak symmetry breaking. This creates a mass splitting between the CP-even and CP-odd sneutrinos. Due to the splitting a particle created as a sneutrino or an antisneutrino evolves towards a state of definite CP, which has no definite lepton number.

In order to have observable lepton number violation, the mass splitting needs to be at least similar in size to the decay width of the sneutrino \cite{Hirsch:1997is,Grossman:1997is}. We explore a case, where the sneutrino decays only through the Yukawa couplings so the decay width is very small.

The scalar potential (\ref{eq:scalarpotential}) also generates a coupling between the Higgs doublets and right-handed sneutrinos. If $\tan \beta$ is large and we are at the alignment limit, $H_{u}$ is close to the SM Higgs and $H_{d}$ to the heavier doublet state. Since the SM-like Higgs has the vacuum expectation value, the heavy Higgs gets the large coupling to sneutrinos. If we assume $\lambda$ and $\lambda_{N}$ to be large and if the decay $H\rightarrow \tilde{N}\tilde{N}$ is allowed, the branching ratio can be fairly large.

\section{Finding sneutrinos at the LHC}

\begin{figure}[hbt]
\begin{center}
\includegraphics[width=0.65\textwidth]{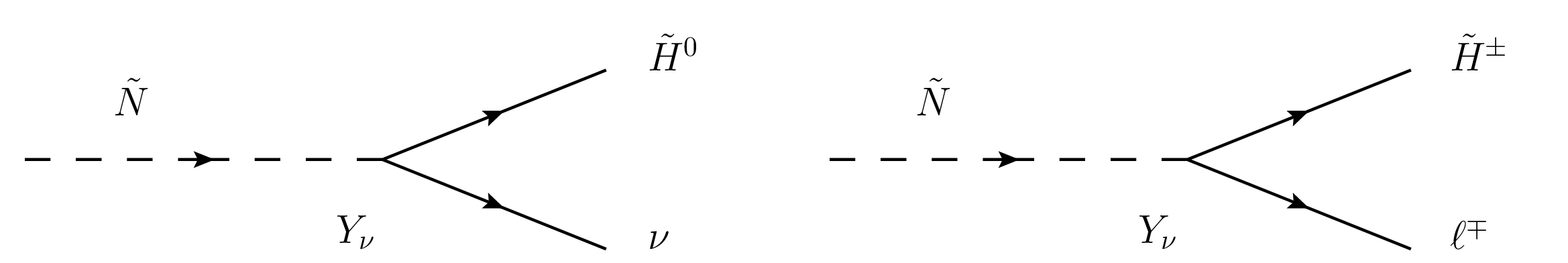}
\end{center}
\caption{The decay modes of the right-handed sneutrino. The amplitudes for both of these processes are proportional to the neutrino Yukawa couplings.\label{fig:snudecay}}
\end{figure}

Our scenario has light higgsinos and they are the only superpartners that are lighter than the right-handed sneutrinos. In such a case the right-handed sneutrinos can decay to a neutrino and a neutralino or a charged lepton and chargino as is shown in figure \ref{fig:snudecay}. Both of these decay amplitudes are proportional to the neutrino Yukawa coupling.

\begin{figure}[hbt]
\begin{center}
\includegraphics[width=0.55\textwidth]{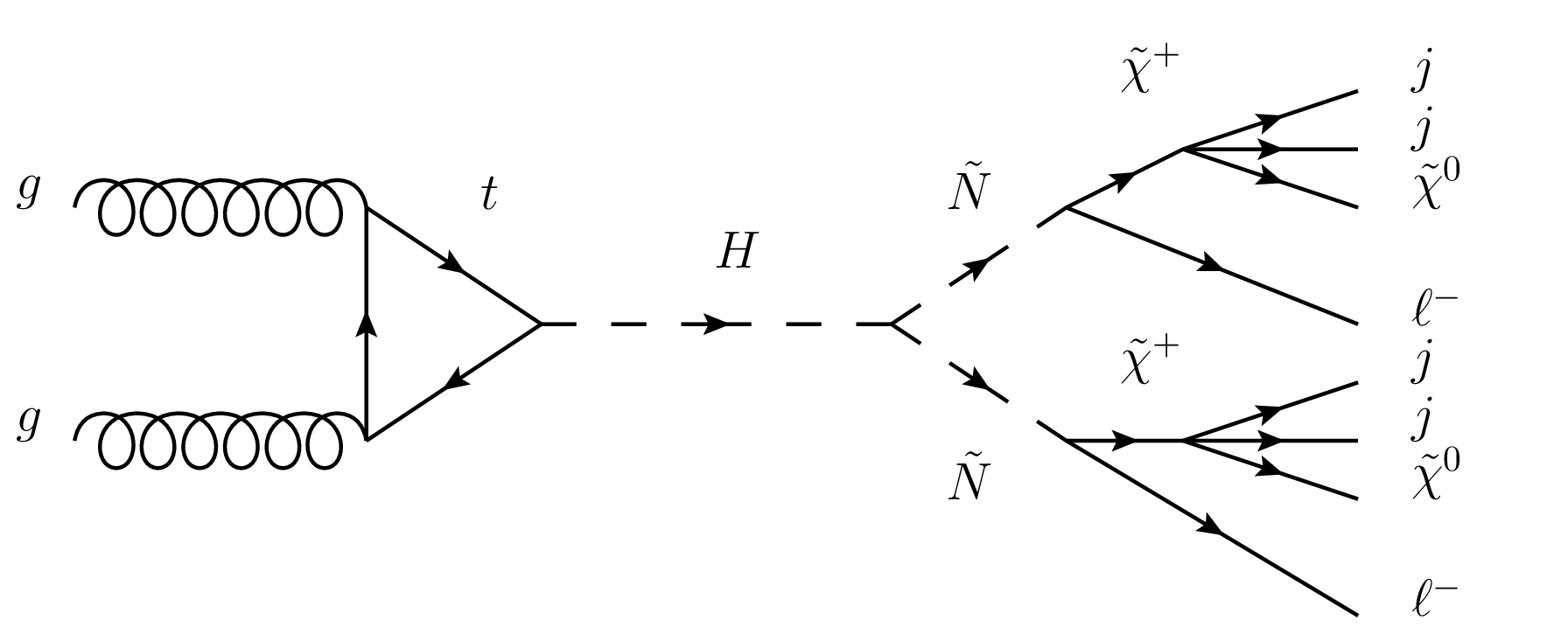}
\end{center}
\caption{The full process we are interested in. There are also similar diagrams for the CP-odd Higgs.\label{fig:snucascade}}
\end{figure}

The overall process is shown in figure \ref{fig:snucascade}. As the sneutrinos have no definite lepton number, they have an equal probability to decay to either sign leptons. We are interested in the same-sign dilepton signature with missing transverse energy from the neutralino (and possible neutrinos).

The main backgrounds come from same-sign $WW$ production, $WZ$ production, where one lepton is missed and from nonprompt leptons, which emerge \textit{e.g.} from $t\overline{t}$ production.

We select events with two same-sign same-flavor leptons, one having $p_{T}> 25$~GeV and the second $p_{T}> 12$~GeV and veto for a third lepton with $p_{T}> 20$~GeV as a third lepton in the signal would emerge from the chargino, which is almost degenerate with the neutralino. In addition we veto for $Z$-bosons (opposite-sign same-flavor leptons with invariant mass $\in [80,100]$~GeV) to suppress the $WZ$ background. We then construct two signal regions, SR1 being better for somewhat compressed spectra and SR2 for larger mass splittings. The cuts are given in table \ref{tb:cuts}.

\begin{table}
\tbl{The cuts defining the two signal regions. The $b$-veto means that there are no objects matching the loose identification criteria of \cite{Chatrchyan:2012jua}. The transverse mass is defined as $M_{T}(\ell_{2}, \slashed{p}_{T})=\sqrt{2\slashed{p}_{T}p_{T}(\ell_{2})(1-\cos(\Delta\phi))}$.\label{tb:cuts}}{
\begin{tabular}{|l|c|c|}
\hline
Cut & SR1 & SR2\\
\hline
Missing transverse energy & & \\
$\slashed{E}_{T}$ & $>50$~GeV & $>100$~GeV\\
\hline
Lepton pair invariant mass & &\\
$M(\ell_{1}\ell_{2})$ & $>10$~GeV & $>10$~GeV\\
& $<50$ GeV & $<80$ GeV \\
\hline
Veto for b-jets: $N(b)$ & $0$ & $0$\\
\hline
Cut on second lepton $M_{T}$ & & \\
$M_{T}(\ell_{2}, \slashed{E}_{T})$ & $> 100$~GeV & $> 100$~GeV\\
\hline
\end{tabular}}
\end{table}

\begin{figure}
\begin{center}
\includegraphics[width=0.45\textwidth]{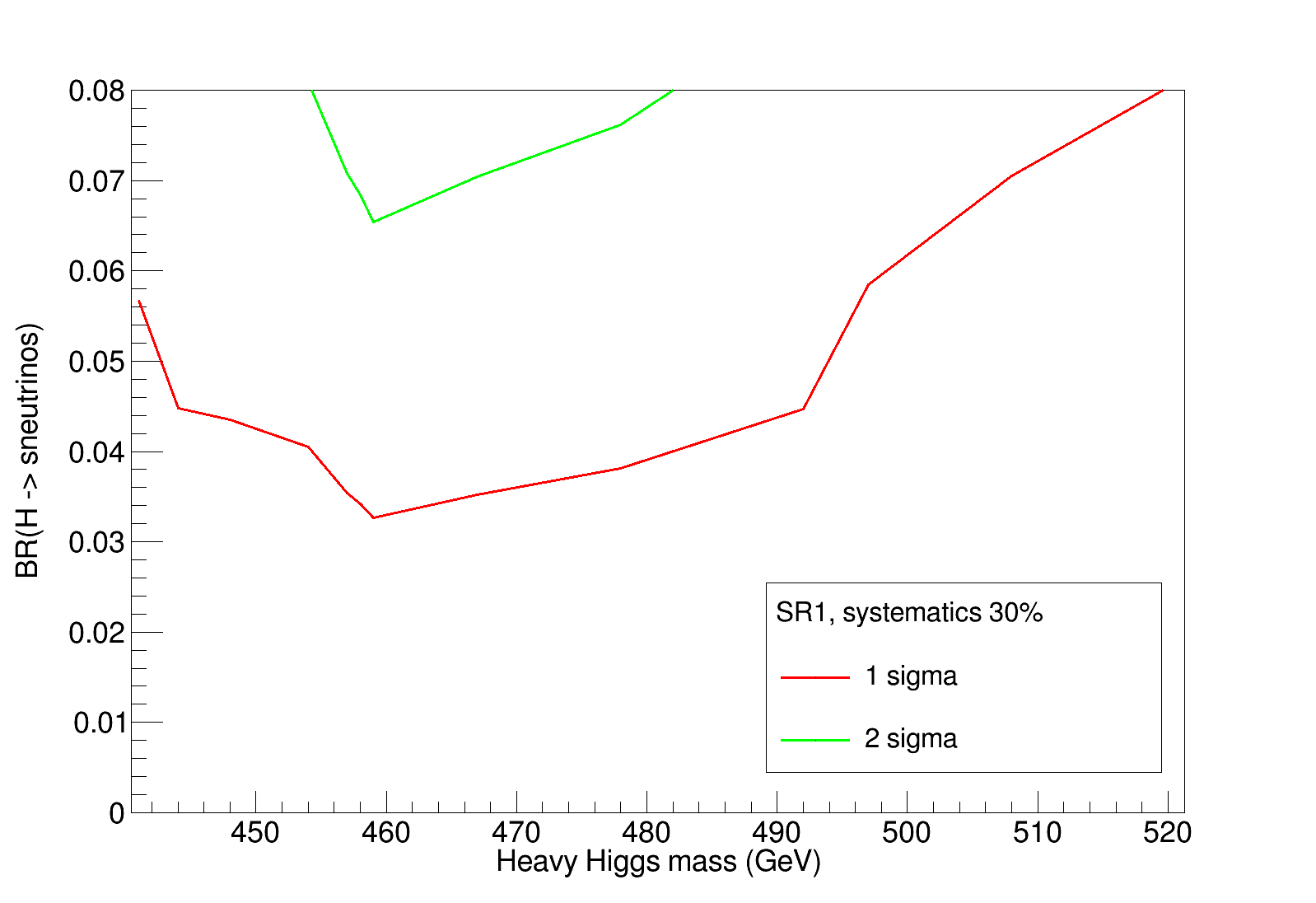}
\includegraphics[width=0.45\textwidth]{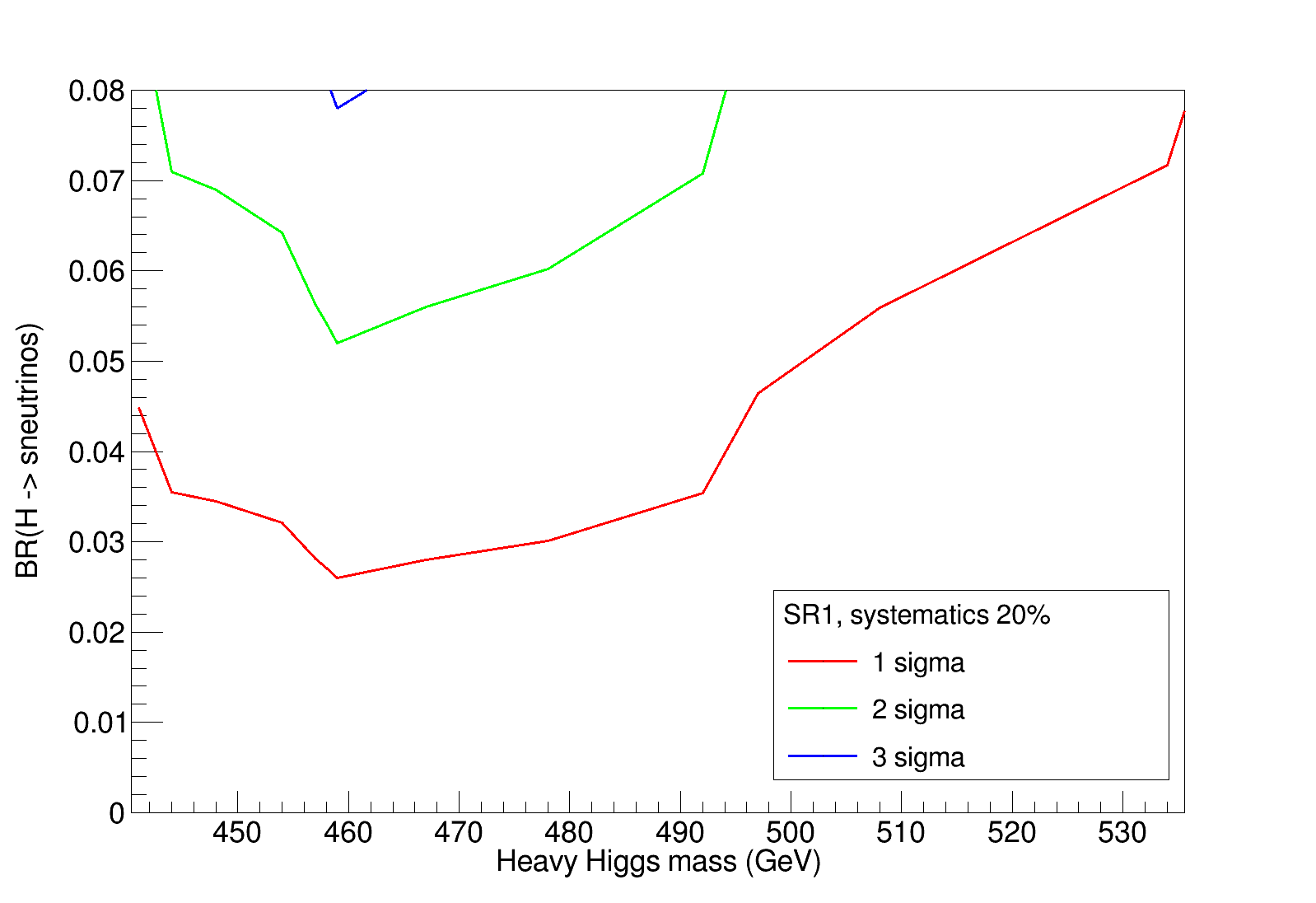}
\includegraphics[width=0.45\textwidth]{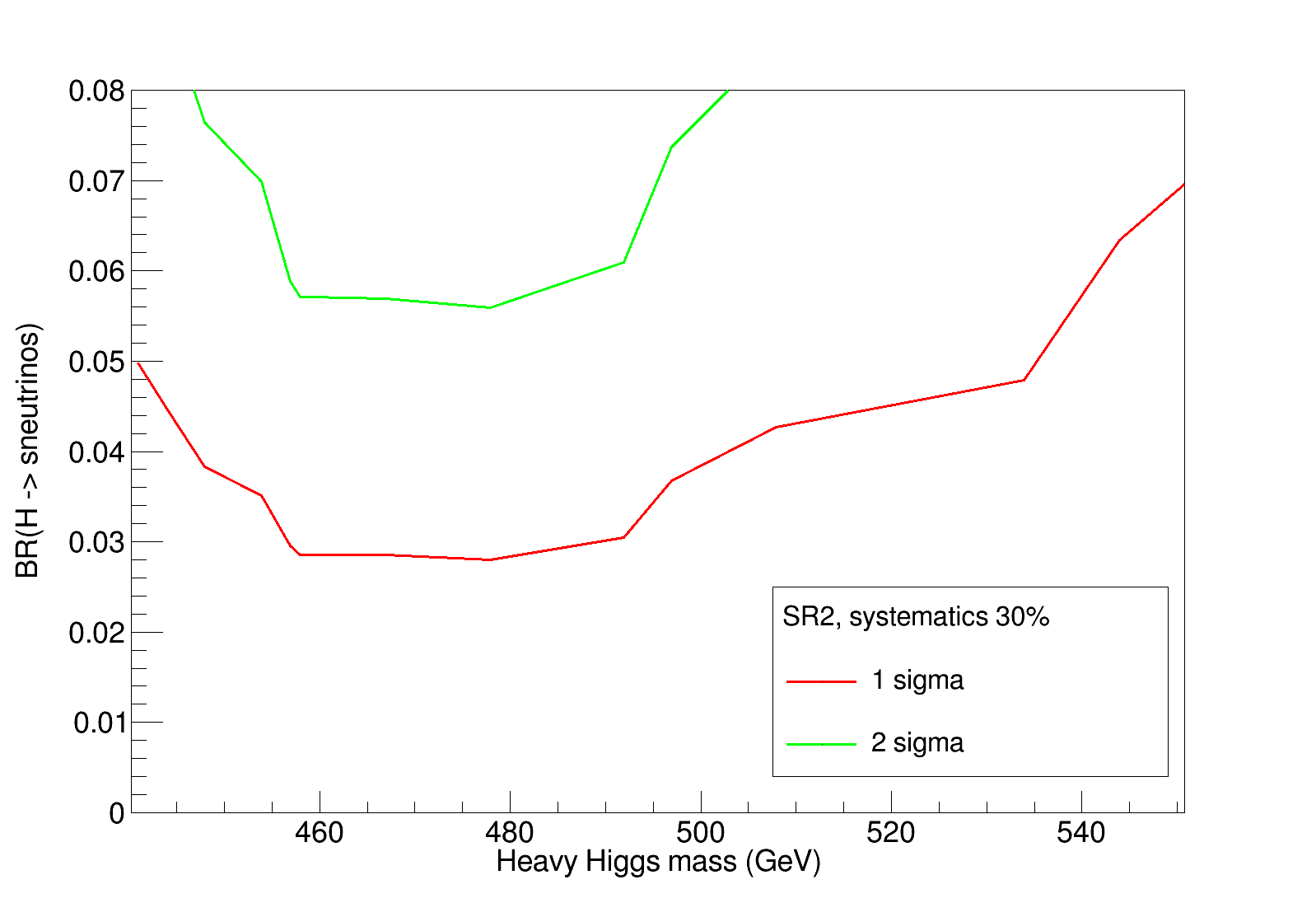}
\includegraphics[width=0.45\textwidth]{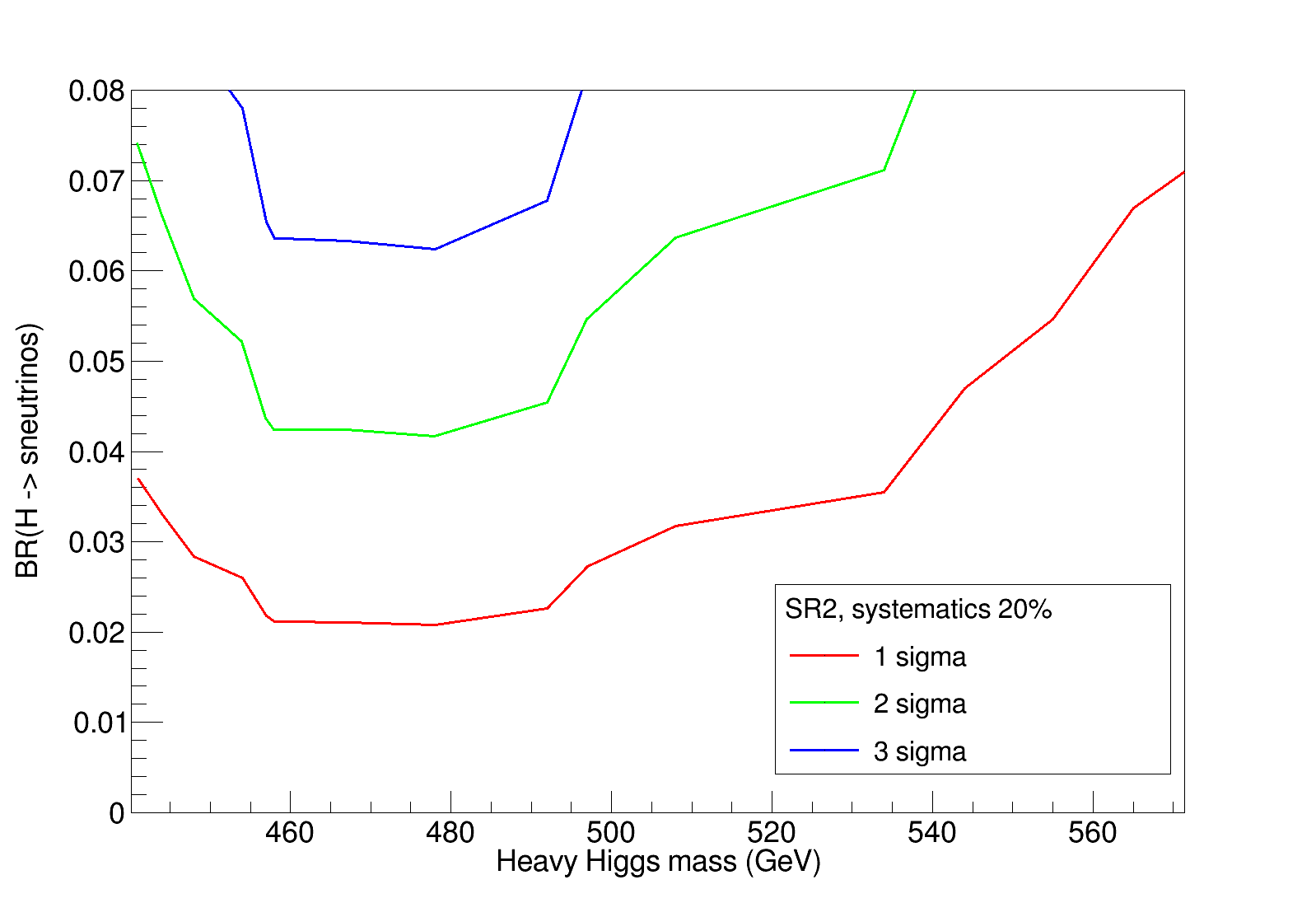}
\end{center}
\caption{The predicted sensitivity of the two signal regions for a benchmark scenario with $m_{\tilde{N}}=220$~GeV and $m_{\tilde{\chi}^{\pm}}=186$~GeV. We have used two assumptions for the total systematic errors, $30\%$ and $20\%$. \label{fig:sensitivityplots}}
\end{figure}

In figure \ref{fig:sensitivityplots} we show the estimated sensitivity for a benchmark scenario with $137$~fb$^{-1}$ of integrated luminosity. We show the sensitivities assuming two baselines for total systematic uncertainties, $30\%$ and $20\%$, the former being close to current uncertainties in similar analyses, the latter could be achieved in the future.

We have some chances of seeing an excess if the branching ratio $H\rightarrow \tilde{N}\tilde{N}$ is large. We probably would not expect the sensitivity to be at a discovery level as it would be quite surprising to discover the heavy Higgs from a sneutrino channel before an excess is seen in one of the SM fermion channels.

\section{Limits on Yukawa couplings}

If we would see a signal of this type at some point, we could use the data to get constraints on the neutrino Yukawa couplings. As the decay amplitude is proportional to the neutrino Yukawa coupling, we get ratios $|y_{ik}/y_{jk}|^{2}$ if we are able to measure the decays in more than one lepton flavor.

An  upper limit on the Yukawa couplings can be obtained from $\sum m_{\nu}=\mathrm{Tr}(m^{\nu})=\sum_{i,j}|y^{\nu}_{ij}|^{2}v^{2}\sin^{2}\beta/2m_{N_{j}}$, but this depends on the value of $\tan\beta$. If we see a sneutrino signal, we would probably see a signal of the heavy Higgs in some channel that gives its mass and if soft SUSY breaking masses are non-negative,  the right-handed neutrino masses are bound by $m_{N}<m_{H}/2$ so we can get an upper bound for the Yukawa coupling from the bound on neutrino masses. A lower bound can be obtained from the decay width if the decays are prompt or from the decay length, if there is a secondary vertex. We assume that also the chargino and neutralino masses can be measured so that the phase space for the sneutrino decays is known.

\section{Conclusions}
Supersymmetry relates the interactions of sneutrinos to those of neutrinos. We showed that in the NMSSM with right-handed neutrinos it is possible to produce sneutrinos through the heavy Higgs portal and the lepton number violating signal could produce an observable excess at the LHC in some corners of the parameter space. If a signal was seen, we could also place some limits on neutrino Yukawa couplings making it possible to study neutrino physics at colliders.

\section*{Acknowledgments}

SM is supported in part through the NExT Institute and the STFC consolidated Grant ST/L000296/1. SM and HW acknowledge the H2020-MSCA-RISE-2014 grant no. 645722 (NonMinimalHiggs). HW acknowledges the support from Magnus Ehrnrooth Foundation and STFC Rutherford International Fellowship (funded through MSCA-COFUND-FP, grant number 665593).

\bibliographystyle{unsrt}

\begin{thebibliography}{99}
%\cite{Athanassopoulos:1997pv}
\bibitem{Athanassopoulos:1997pv}
  C.~Athanassopoulos {\it et al.} [LSND Collaboration],
  %``Evidence for nu(mu) ---> nu(e) neutrino oscillations from LSND,''
  Phys.\ Rev.\ Lett.\  {\bf 81} (1998) 1774
 % %doi:10.1103/PhysRevLett.81.1774
  [nucl-ex/9709006].
  %%CITATION = %doi:10.1103/PhysRevLett.81.1774;%%

%\cite{Fukuda:1998mi}
\bibitem{Fukuda:1998mi}
  Y.~Fukuda {\it et al.} [Super-Kamiokande Collaboration],
  %``Evidence for oscillation of atmospheric neutrinos,''
  Phys.\ Rev.\ Lett.\  {\bf 81} (1998) 1562
  %%doi:10.1103/PhysRevLett.81.1562
  [hep-ex/9807003].
  %%CITATION = %doi:10.1103/PhysRevLett.81.1562;%%

%\cite{Aker:2019uuj}
\bibitem{Aker:2019uuj}
  M.~Aker {\it et al.} [KATRIN Collaboration],
  %``An improved upper limit on the neutrino mass from a direct kinematic method by KATRIN,''
  Phys.\ Rev.\ Lett.\  {\bf 123} (2019) no.22,  221802
%  doi:10.1103/PhysRevLett.123.221802
  [arXiv:1909.06048 [hep-ex]].
  %%CITATION = doi:10.1103/PhysRevLett.123.221802;%%

%\cite{Aghanim:2018eyx}
\bibitem{Aghanim:2018eyx}
  N.~Aghanim {\it et al.} [Planck Collaboration],
  %``Planck 2018 results. VI. Cosmological parameters,''
  arXiv:1807.06209 [astro-ph.CO].
  %%CITATION = ARXIV:1807.06209;%%

%\cite{Weinberg:1979sa}
\bibitem{Weinberg:1979sa}
  S.~Weinberg,
  %``Baryon and Lepton Nonconserving Processes,''
  Phys.\ Rev.\ Lett.\  {\bf 43} (1979) 1566.
%  doi:10.1103/PhysRevLett.43.1566
  %%CITATION = doi:10.1103/PhysRevLett.43.1566;%%

%\cite{Moretti:2019yln}
\bibitem{Moretti:2019yln}
  S.~Moretti, C.~Shepherd-Themistocleous and H.~Waltari,
  %``Lepton number violation in heavy Higgs decays to sneutrinos,''
  Phys.\ Rev.\ D {\bf 101} (2020) no.1,  015018
%  doi:10.1103/PhysRevD.101.015018
  [arXiv:1909.04692 [hep-ph]].
  %%CITATION = doi:10.1103/PhysRevD.101.015018;%%

%\cite{Kitano:1999qb}
\bibitem{Kitano:1999qb}
  R.~Kitano and K.~y.~Oda,
  %``Neutrino masses in the supersymmetric standard model with right-handed neutrinos and spontaneous R-parity violation,''
  Phys.\ Rev.\ D {\bf 61} (2000) 113001
 % %doi:10.1103/PhysRevD.61.113001
  [hep-ph/9911327].
  %%CITATION = %doi:10.1103/PhysRevD.61.113001;%%

%\cite{Cerdeno:2008ep}
\bibitem{Cerdeno:2008ep}
  D.~G.~Cerdeno, C.~Munoz and O.~Seto,
  %``Right-handed sneutrino as thermal dark matter,''
  Phys.\ Rev.\ D {\bf 79} (2009) 023510
 % %doi:10.1103/PhysRevD.79.023510
  [arXiv:0807.3029 [hep-ph]].
  %%CITATION = %doi:10.1103/PhysRevD.79.023510;%%

%\cite{Hirsch:1997is}
\bibitem{Hirsch:1997is}
  M.~Hirsch, H.~V.~Klapdor-Kleingrothaus and S.~G.~Kovalenko,
  %``Sneutrino oscillations and neutrinoless double beta decay,''
  Phys.\ Lett.\ B {\bf 403} (1997) 291.
%  doi:10.1016/S0370-2693(97)00481-4
  %%CITATION = doi:10.1016/S0370-2693(97)00481-4;%%

%\cite{Grossman:1997is}
\bibitem{Grossman:1997is}
  Y.~Grossman and H.~E.~Haber,
  %``Sneutrino mixing phenomena,''
  Phys.\ Rev.\ Lett.\  {\bf 78} (1997) 3438
%  doi:10.1103/PhysRevLett.78.3438
  [hep-ph/9702421].
  %%CITATION = doi:10.1103/PhysRevLett.78.3438;%%

%\cite{Chatrchyan:2012jua}
\bibitem{Chatrchyan:2012jua}
  S.~Chatrchyan {\it et al.} [CMS Collaboration],
  %``Identification of b-Quark Jets with the CMS Experiment,''
  JINST {\bf 8} (2013) P04013
%  doi:10.1088/1748-0221/8/04/P04013
  [arXiv:1211.4462 [hep-ex]].
  %%CITATION = doi:10.1088/1748-0221/8/04/P04013;%%

\end{thebibliography}

\end{document}